\documentclass[aps,twocolumn,pra]{revtex4}
\usepackage{epsfig,rotating,minitoc}
\usepackage{amsmath,amssymb,amsthm}
\usepackage{graphicx}
\usepackage{psfrag}
\usepackage{bm}
\usepackage[dvips]{color}

\begin{document}

\title{Husimi distribution and phase space analysis of Dicke model quantum phase transition} 

\author{E. Romera}
\affiliation{Departamento de F\'{\i}sica At\'omica, Molecular y Nuclear and
Instituto Carlos I de F{\'\i}sica Te\'orica y
Computacional, Universidad de Granada, Fuentenueva s/n, 18071 Granada,
Spain}
\author{R. del Real}
\affiliation{Departamento de F\'{\i}sica At\'omica, Molecular y Nuclear, Universidad de Granada, Fuentenueva s/n, 18071 Granada,
Spain}
\author{M. Calixto}
\affiliation{Departamento de Matem\'atica Aplicada, Universidad de Granada,
Fuentenueva s/n, 18071 Granada, Spain}

\date{\today}
\begin{abstract}
The Husimi distribution is proposed for a phase space analysis of quantum phase transitions
in the Dicke model of spin-boson interactions. We show that the  
inverse participation ratio and Wehrl entropy of the Husimi distribution give sharp signatures of the quantum phase transition. 
The analysis has been done using two frameworks: a numerical
treatment and  an analytical  variational approximation.  Additionally we have
proposed a new characterization of 
  the Dicke model  quantum phase transition 
by means of the zeros of the Husimi distribution in the variational
approach.

\end{abstract}
\maketitle

\section{Introduction}

Understanding quantum phase transitions (QPTs) is a relevant fact in quantum many
body problems \cite{sachdev}. QPTs usually occur in a system described by a 
Hamiltonian of the form $H=H_0+\lambda H_1$, with $H_0$ exactly solvable, $H_1$ an interaction term and $\lambda$ 
the corresponding control interaction-strength parameter. The 
QPT occurs when $\lambda$ reaches some critical value $\lambda_c$ in which the
\textcolor{black}{symmetries  (and consequently the properties)} of the
system change drastically. This is the case of an ensemble of atoms interacting with a single bosonic field
mode described by the Dicke Hamiltonian.

Quantum mechanics offers different distributions to characterize
phase-space properties \cite{Gerry2005}. One is the Wigner function, widely used in quantum optics. Other is the 
Husimi distribution,  which  is given by the
overlap between a minimal uncertainty (coherent) state and the wavefunction. 
This distribution is sometimes more convenient because, unlike Wigner distribution, 
it is non-negative. Husimi distribution has been found useful for a
phase-space visualization of a metal-insulator transition \cite{aulbach}, to analyze
quantum chaos in atomic physics \cite{monteiro} or to analyze models in condensed
matter physics \cite{weinmann99}. In addition, we would like to point out that the zeros of the Husimi distribution 
have essential information, in particular, the quantum state can be  described by its distribution of
zeros \cite{Korsch}. They are simply
the least probable points in phase space and they have been considered as a
quantum indicator of classical and quantum chaos \cite{4,dando}

The Husimi distribution has a great amount of information and it can be useful
to consider informational measures as the so-called inverse participation ratio and Wehrl entropy
\cite{Mintert}. Recently, an analysis of QPT in the Dicke
model by means of information measures \cite{pla11,jsm11,epl2012,physica12,renyipra} has been done in position and momentum
spaces, separately. Here we will do an informational description of the Dicke model QPT in 
phase space in terms of the inverse participation
ration  (and higher moments) and the Wehrl entropy of the Husimi distribution and its marginals. Additionally we will
investigate the visualization of Dicke model QPT through the zeros of the Husimi
distribution.
 
This article is organized as follows. In Section \ref{sec1} we briefly remind the Dicke Hamiltonian, introduce coherent states and the Husimi distribution of 
the ground state,  define moments, R\'ennyi-Wehrl entropies and 
marginals of the Husimi distribution and 
present numerical results. In Section \ref{sec2} we study  a variational approximation to the ground state wave function 
in terms of symmetry-adapted coherent states and analyze the information measures in the thermodynamic limit. Zeros of the Husimi (ansatz) distribution are also 
computed and graphically represented in order to characterize the QPT. 


\section{Dicke Hamiltonian and Husimi distribution}\label{sec1}

The single-mode  Dicke model is a well studied object in the field of QPTs \cite{emaryprl2,emarypra,brandes}. In this case the Hamiltonian is given by 
\begin{equation}
\label{qpt01}
H=\omega_0 J_z + \omega a^{\dag} a + \frac{\lambda}{\sqrt{2 j}}
( a^{\dag} + a )( J_+ + J_-) ,
\end{equation}
describing an ensemble of $N$ two-level atoms with level-splitting $\omega_0$, with 
$J_z$, $J_{\pm}$ the  angular momentum operators for a
pseudospin of length $j=N/2$, and $a$ and $a^{\dag}$ are the 
bosonic operators of
the field with frequency $\omega$. It is well known that
there is a QPT at the critical value of the coupling parameter
$\lambda=\lambda_c=\frac{\sqrt{\omega\omega_0}}{2}$ from 
the so-called normal phase ($\lambda<\lambda_c$)  to the superradiant phase
($\lambda > \lambda_c$). 
Let us consider a basis set $\left\{|n;j,m\rangle\equiv |n\rangle\otimes|j,m\rangle \right\}$ of
the Hilbert space, with $\left\{|n\rangle\right\}_{n=0}^{\infty}$ the number
states of the field and $\left\{|j,m\rangle\right\}_{m=-j}^{j}$ the so called
Dicke states of the atomic sector. The matrix elements of the Hamiltonian in this basis are:
\begin{eqnarray}
\langle n^{\prime};
j^{\prime},m^{\prime}|H|n;j,m\rangle= (n\omega+m\omega_0)
\delta_{n^{\prime},n}\delta_{m^{\prime},m}
\notag\\ 
+\frac{\lambda}{\sqrt{2j}}(\sqrt{n+1}\delta_{n^{\prime},n+1}+\sqrt{n}\delta_{n^{\prime},n-1})\notag \\ \times (\sqrt{j(j+1)-m(m+1)}
\delta_{m^{\prime},m+1}\notag\\ +\sqrt{j(j+1)-m(m-1)}\delta_{m^{\prime},m-1}).\label{hamel}
\end{eqnarray}
At this point it is important to note that time evolution preserves the parity $e^{i\pi(n+m+j)}$ of a given state 
$|n;j,m\rangle$. That is, the parity operator $\hat\Pi=e^{i\pi(a^\dag a+J_z+j)}$ commutes with ${H}$ and both operators 
can then be jointly diagonalized. In particular, the ground state must be even (see later on Eq. (\ref{sacs})).

\subsection{Coherent states and Husimi distribution}

Let us denote by 
\begin{equation}
\begin{array}{l}
|\alpha\rangle=e^{-|\alpha|^2/2}e^{\alpha a^\dag}|0\rangle=
e^{-|\alpha|^2/2}\sum_{n=0}^\infty\frac{\alpha^n}{\sqrt{n!}}|n\rangle,\\
|z\rangle=(1+|z|^2)^{-j}e^{zJ_+}|j,-j\rangle=\\(1+|z|^2)^{-j}\sum_{m=-j}^j\binom{2j}{j+m}^{1/2}z^{j+m}|j,m\rangle,
\end{array}\label{cohs}
\end{equation}
(with $\alpha,z\in\mathbb C$) the  standard (canonical or Glauber) and spin-$j$ Coherent States (CSs) for the photon and the
particle sectors, respectively. It is well known (see e.g. \cite{Perelomov}) that  coherent states form an overcomplete set 
of the corresponding Hilbert space and fulfill 
the closure relations or resolutions of the identity:
\begin{eqnarray}
1&=&\frac{1}{\pi}\int_{{\mathbb R}^2}|\alpha\rangle\langle\alpha|d^2\alpha, \nonumber\\ 
1&=&\frac{2j+1}{\pi}\int_{{\mathbb R}^2}|z\rangle\langle z|\frac{d^2 z}{(1+|z|^2)^2},\label{closure}
\end{eqnarray}
with $d^2w=d\mathrm{Re}(w)=d\mathrm{Im}(w)$ the Lebesgue measure on $\mathbb R^2$ or $\mathbb C$. It is 
straightforward to see that the probability amplitude of detecting $n$ photons and $j+m$ excited atoms in 
$|\alpha,z\rangle\equiv |\alpha\rangle\otimes|z\rangle$ is given by:
\begin{equation}
 \varphi_{n,m}^{(j)}(\alpha,z)=\langle n|\alpha\rangle\langle j,m|z\rangle=\frac{e^{-|\alpha|^2/2}\alpha^n}{\sqrt{n!}}
\frac{\sqrt{\binom{2j}{j+m}}z^{j+m}}{(1+|z|^2)^j}.
\end{equation}
The ground state vector $\psi$ will be given as an expansion 
\begin{equation}
 |\psi\rangle=\sum_{n=0}^{n_c}\sum_{m=-j}^{j} c_{nm}^{(j)}|n;j,m\rangle
\end{equation}
where the coefficients $c_{nm}^{(j)}$ are calculated by numerical diagonalization of \eqref{hamel} with a given cutoff $n_c$. The 
Husimi distribution of $\psi$ is then given by
\begin{eqnarray}
 {\Psi}(\alpha,z)&=&|\langle\alpha,z|\psi\rangle|^2\label{husiz}\\ &=&\sum_{n,n'=0}^{n_c}\sum_{m,m'=-j}^{j} c_{nm}^{(j)} 
\bar c_{n'm'}^{(j)}\varphi_{n,m}^{(j)}(\alpha,z)\varphi_{n,m}^{(j)}(\bar\alpha,\bar z)\nonumber
\end{eqnarray}
and normalized according to:
\begin{equation}
\frac{2j+1}{\pi^2} \int_{\mathbb R^4} {\Psi}(\alpha,z){d^2\alpha}\frac{d^2 z}{(1+|z|^2)^2}=1.
\end{equation}
Before discussing marginals of the Husimi distribution and their properties, let us introduce an important 
approximation which will simplify things greatly.

\subsection{Holstein-Primakoff representation and large pseudospin}

We shall make use of the Holstein-Primakoff representation \cite{HP} of the angular momentum operators 
$J_\pm, J_z$ in terms of the bosonic operators, $[b,b^\dag]=1$, given by:
\begin{equation}
\begin{array}{c}
J_+=b^\dag\sqrt{2j-b^\dag b}, \;\;\; J_-=\sqrt{2j-b^\dag b}\;b
\\ J_z=(b^\dag b-j).
\end{array}\label{hprep}
\end{equation}
For high values of $j$ (and fixed $b^\dag b$), we can approximate $J_+\simeq \sqrt{2j}\,b^\dag$ and 
$J_-\simeq \sqrt{2j}\,b$, so that the atomic sector 
can be practically described by an harmonic oscillator, just like the field sector. 
Introducing then position and momentum operators for the two bosonic modes as usual:
\begin{equation}
\begin{array}{ll} X=\frac{1}{\sqrt{2\omega}}(a^\dag+a), & P_X=i\sqrt{\frac{\omega}{2}}(
a^\dag-a),\\
Y=\frac{1}{\sqrt{2\omega_0}}(b^\dag+b), &P_Y=i\sqrt{\frac{\omega_0}{2}}(b^\dag+b),\end{array}\label{posmomenop}
\end{equation}
the wave function position  representation is formally
equivalent to that of a set of two coupled harmonic oscillators and can be written as
\cite{emarypre}:
\begin{equation}
\psi(x,y)=\sqrt{\omega\omega_0} e^{-\frac{1}{2}(\omega x^2+ \omega_0
y^2)}\sum_{n=0}^{n_c}\sum_{m=-j}^{j} c_{nm}^{(j)}
\notag
\end{equation}
\begin{equation}
\times
\frac{H_n(\sqrt{\omega}x)H_{j+m}(\sqrt{\omega_0}y)}{2^{(n+m+j)/2}\sqrt{n!(j+m)!}}
\label{wf}
\end{equation}
where we have made use of the definition of
\begin{eqnarray}
 \langle x| n\rangle &=&   \sqrt{\omega} e^{-\frac{1}{2}\omega x^2}\frac{
   H_n(\sqrt{\omega} x)}{\sqrt{2^n
       n!\sqrt{\pi}}},\\ \langle y| j,m\rangle &=& \sqrt{\omega_0}e^{-\frac{1}{2}\omega_0 y^2} \frac{H_{j+m}(\sqrt{\omega_0} y)}{\sqrt{2^{(j+m)}
       (j+m)!\sqrt{\pi}}},\nonumber 
\end{eqnarray}
(the Hermite polynomials of degree $n$ and $j+m$, respectively) and we have truncated the Hilbert space of the 
field sector to dimension $n_c$ looking for the numerical solution and convergence of the eigenproblem \cite{Hirchdetallado}. 
This is a very convenient representation that 
has already been used in Ref.  \cite{emarypre}. Analogously in momentum space:
 \begin{equation}
\tilde{\psi}(p_x,p_y)=\frac{1}{\sqrt{\omega\omega_0}} e^{-\frac{1}{2}( \frac{p_x^2}{\omega}+ \frac{
p_y^2}{\omega_0})}\sum_{n=0}^{n_c}\sum_{m=-j}^{j} (-i)^{n+m+j} c_{nm}^{(j)}
\notag
\end{equation}
\begin{equation}
\times
\frac{H_n(p_x/\sqrt{\omega})H_{j+m}(p_y/\sqrt{\omega_0})}{2^{(n+m+j)/2}\sqrt{n!(j+m)!}}.
\label{wfmomentum}
\end{equation}
Moreover, redefining 
\begin{equation}
 \beta\equiv \sqrt{2j}\,z,\label{hp1}
\end{equation}
in \eqref{cohs}, it can be seen (see e.g. \cite{CScontract1,Perelomov}) that 
spin-$j$ coherent states $|z\rangle$ go over to ordinary coherent states
  $|\beta\rangle\equiv e^{-|\beta|^2/2}e^{\beta b^\dag}|0\rangle$  for $j\gg 1$ 
(when identifying  $|j,-j\rangle\equiv|0\rangle$ and $|j,m\rangle\equiv|m+j\rangle$). Thus, we shall assume the approximation:
\begin{equation}
|z\rangle\simeq|\beta\rangle,\label{contract}
\end{equation}
which turns out to be a quite good estimate, even for relatively small values of $j$, for $|z|$ in a neighborhood of the equilibrium 
value $|z_e|<1$  in \eqref{critpoints}. With this approximation, the 
Husimi distribution \eqref{husiz} becomes
\begin{eqnarray}
 \Phi(\alpha,\beta)&=&|\langle\alpha,\beta|\psi\rangle|^2\label{husibeta}\\ &=&\sum_{n,n'=0}^{n_c}\sum_{m,m'=-j}^{j} c_{nm}^{(j)} 
\bar c_{n'm'}^{(j)}\phi_{n,m}^{(j)}(\alpha,\beta)\phi_{n',m'}^{(j)}(\bar\alpha,\bar \beta),\nonumber
\end{eqnarray}
where now
\begin{equation}
\phi_{n,m}^{(j)}(\alpha,\beta)=\langle n|\alpha\rangle\langle j+m|\beta\rangle=\frac{e^{-|\alpha|^2/2}\alpha^n}{\sqrt{n!}}
\frac{e^{-|\beta|^2/2}\beta^{m+j}}{\sqrt{(m+j)!}}
\end{equation}
with the new normalization
\begin{equation}
 \int_{{\mathbb R}^4} \Phi(\alpha,\beta)\frac{d^2\alpha d^2\beta}{\pi^2}=1.\label{Qnorm}
\end{equation}

\subsection{Moments, R\'enyi-Wehrl entropy and marginals of the Husimi distribution}

Important quantities to visualize the QPT in the Dicke model across the critical point $\lambda_c$ will be the 
$\nu$-th moments of the Husimi distribution \eqref{husibeta}:
\begin{equation}
 M_{j,\nu}(\lambda)=\int_{{\mathbb R}^4}\frac{d^2\alpha d^2\beta}{\pi^2} (\Phi(\alpha,\beta))^\nu.\label{momentsnu}
\end{equation}
Note that $M_{j,1}=1$ since $\Phi$ is normalized \eqref{Qnorm}. Among all moments we shall single-out 
the so-called ``inverse participation ratio'' $P_j(\lambda)=M_{j,2}(\lambda)$ which somehow measures the (de-)localization 
of $\Phi$ across the phase transition. The definition of the moments $M_{j,\nu}$ is not restricted to integer values of $\nu$.  
Once $M_{j,\nu}$ are known for all integer $\nu$, there is a unique
analytic extension to complex (and therefore real) $\nu$, as
integers are dense at infinity. The ``classical'' (versus quantum von Neumann) R\'enyi-Wehrl entropy is then defined as:
\begin{equation}
W_{j,\nu}(\lambda)=\frac{1}{1-\nu}\ln(M_{j,\nu}(\lambda)),\label{wehrlnu}
\end{equation}
which tends to the  Wehrl entropy
\begin{equation}
W_{j}(\lambda)=-\int_{{\mathbb R}^4}\frac{d^2\alpha d^2\beta}{\pi^2} \,\, \Phi(\alpha,\beta)\ln \Phi(\alpha,\beta)\label{wehrl1}
\end{equation}
when $\nu\to 1$.

In order to differentiate between position and momentum behaviors, we shall study the marginals 
of the Husimi distribution in each space:
\begin{eqnarray}
 \Phi_1(\alpha_1,\beta_1)&=&\int_{{\mathbb R}^2}\frac{d\alpha_2 d\beta_2}{\pi} \Phi(\alpha_1+i\alpha_2,\beta_1+i\beta_2)\nonumber\\
 \Phi_2(\alpha_2,\beta_2)&=&\int_{{\mathbb R}^2}\frac{d\alpha_1 d\beta_1}{\pi} \Phi(\alpha_1+i\alpha_2;\beta_1+i\beta_2),\label{marginalQ}
\end{eqnarray}
so that
\begin{equation}
\int_{{\mathbb R}^2}\frac{d\alpha_{\kappa} d\beta_{\kappa}}{\pi} \Phi_{\kappa}(\alpha_{\kappa},\beta_\kappa)=1,\,\,\kappa=1,2.
\end{equation}
We shall also be interested in the moments of marginal distributions 
\begin{eqnarray}
M_{j,\nu}^{(\kappa)}(\lambda)=\int_{{\mathbb R}^2}\frac{d\alpha_\kappa d\beta_\kappa}{\pi}(\Phi_\kappa(\alpha_\kappa,\beta_\kappa))^\nu,
\,\kappa=1,2,\label{momentsnumarg}
\end{eqnarray}
specially the marginal inverse participation ratios $P_j^{(\kappa)}(\lambda)=M_{j,2}^{(\kappa)}$ and marginal Wehrl entropies 
\begin{equation}
W_j^{(\kappa)}(\lambda)=-\int_{{\mathbb R}^2}\frac{d\alpha_\kappa d\beta_\kappa}{\pi} \,\, \Phi_\kappa(\alpha_\kappa,\beta_\kappa)
\ln \Phi_\kappa(\alpha_\kappa,\beta_\kappa)
\end{equation}
in position ($\kappa=1$) and momentum ($\kappa=2$) spaces 
as a function of the control parameter $\lambda$. In general, $P_j(\lambda)\not= P_j^{(1)}(\lambda)P_j^{(2)}(\lambda)$ and 
$W_j(\lambda)\not=W_j^{(1)}(\lambda)+W_j^{(2)}(\lambda)$, but 
these quantities are approximately equal for high $j$ 
except in a close neighborhood of $\lambda_c$ \cite{aulbach}.

In Appendix \ref{apA} we provide a connection with other phase-space formulas for marginal distributions  in 
position and momentum coordinates \eqref{posmomenop}. This approach has been used in  \cite{aulbach} to  visualize 
the metal-insulator QPT described by the Aubry-Andr\'e model. We also make use of this representation to make numerical 
calculations more maneuverable.

\subsection{Numerical Results}

\begin{figure}
\begin{center}
\includegraphics[angle=-90,width=9cm]{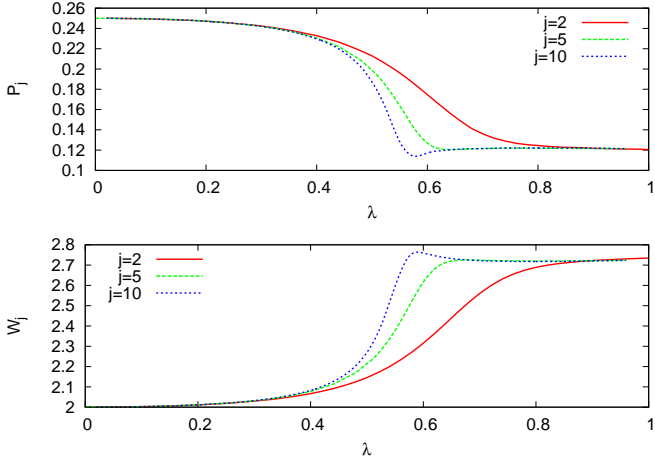}
\end{center}
\caption{Inverse participation ratio $P_j(\lambda)$ and entropy in phase space $W_j(\lambda)$ for
  $j=5$ and $j=10$ and $\omega_0=\omega=1$ as a function of $\lambda$ }
\label{parti}
\end{figure}

\begin{figure}
\begin{center}
\includegraphics[angle=-90,width=9cm]{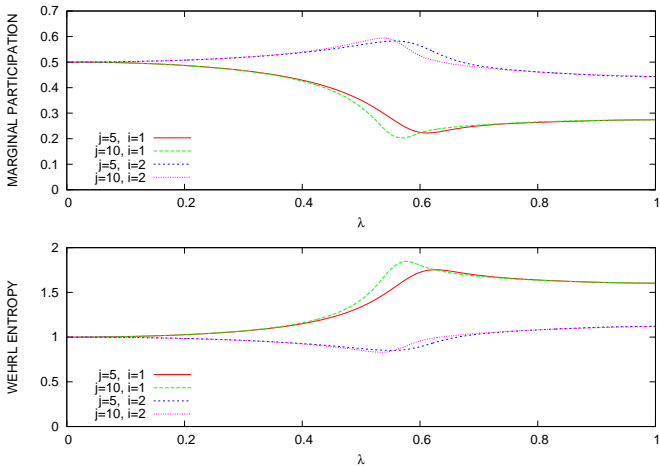}
\end{center}
\caption{Inverse participation ratio $P_j^{(i)}\lambda)$ and entropy in phase space $W_j^{(i)}(\lambda)$ for
   $j=5$ and $j=10$, $i=1,$ $2$ and $\omega_0=\omega=1$ as a function of $\lambda$ }
\label{partimar}
\end{figure}

Firstly we have calculated the participation ratio $P_j(\lambda)$ and the Wherl
entropy $W_j(\lambda)$ for different values of $\lambda$. The computed results
are given in Fig. \ref{parti} where we present $P_j(\lambda)$    and $W_j(\lambda)$ for
$j=2,5,$ $10$ and for $\omega=\omega_0=1$ (for which $\lambda_c=0.5$). Notice that
the inverse partipation ratio (top pannel) is around $1/4$ in the normal phase
decreasing around the critical point to reach the value $1/8$ in the
superradiant phase. We can see that  the change in the
participation ratio is suddener as $j$ increases. The Wherl entropy (botton
panel) is approximately $2$  in the normal phase and around $2+\ln 2$ in the
superradiant phase changing suddenly (suddener as $j$ increases) around the
critical point. \textcolor{black}{For completeness we have represented the computed marginal
quantities  in figure \ref{partimar}.}



\section{Variational aproximation and the thermodynamic limit}\label{sec2}

Now we present analytical expressions for Husimi's distribution, its marginals, moments and entropies using
trial states expressed in terms of ``parity-symmetry-adapted'' CSs introduced by
Casta\~nos et al. \cite{casta1,casta2}, which turn out to be an excellent approximation to the exact quantum solution of
 the ground  (+) and first excited (--) states of the Dicke model. 

\subsection{Symmetry-adapted coherent states and their Husimi distribution}

Using the direct product $|\alpha,z\rangle\equiv |\alpha\rangle\otimes|z\rangle$ as 
a ground-state ansatz, one can easily compute the mean energy 
\begin{equation} \begin{array}{l}{\cal H}(\alpha,z)=\langle \alpha, z|H|\alpha,z\rangle\\
                  =\omega|\alpha|^2+j\omega_0\frac{|z|^2-1}{|z|^2+1}+{\lambda}{\sqrt{2j}}(\alpha+\bar\alpha)\frac{\bar z+z}{|z|^2+1},\\
                 \end{array}
\end{equation}
which defines a four-dimensional ``energy surface''. Minimizing with respect to these 
four coordinates gives the equilibrium points:
\begin{eqnarray}
\alpha_e&=&\left\{\begin{array}{ll} 0, & \mathrm{if}\, \lambda<\lambda_c\\
-\sqrt{2j}\sqrt{\frac{\omega_0}{\omega}}\frac{\lambda}{\lambda_c}\sqrt{1-\left(\frac{\lambda}{\lambda_c}\right)^{-4}}, &
\mathrm{if}\, \lambda\geq\lambda_c\end{array}\right.\nonumber\\
z_e&=&\left\{\begin{array}{ll} 0, & \mathrm{if}\, \lambda<\lambda_c\\
\sqrt{\frac{\frac{\lambda}{\lambda_c}-\left(\frac{\lambda}{\lambda_c}\right)^{-1}}{\frac{\lambda}{\lambda_c}+
\left(\frac{\lambda}{\lambda_c}\right)^{-1}}},
&
\mathrm{if}\, \lambda\geq\lambda_c\end{array}\right.\label{critpoints}
\end{eqnarray}
Note that $\alpha_e$ and $z_e$ are real and non-zero above the critical point $\lambda_c$ (i.e., in the superradiant phase).

Although the direct product  $|\alpha,z\rangle$ gives a good variational approximation to the ground state mean energy in the 
thermodynamic limit $j\to\infty$, 
it does not capture the correct behavior for other ground state 
properties sensitive to the parity symmetry $\hat\Pi$ of the Hamiltonian (\ref{qpt01}) like, for instance, uncertainty and entropy measures.
This is why parity-symmetry adapted coherent states are introduced. Indeed, a far better variational description 
of the ground (resp. first-excited) state is given in terms of the even-(resp. odd)-parity coherent states
\begin{equation}
|\psi_\pm\rangle=|\alpha,z,\pm\rangle=\frac{|\alpha\rangle\otimes|z\rangle\pm|-\alpha\rangle\otimes|-z\rangle}{{\mathcal N}_\pm(\alpha,z)},\label{sacs}
\end{equation}
obtained by applying projectors of even and odd parity $\hat{\cal P}_\pm=(1\pm \hat\Pi)$ to the direct product $|\alpha\rangle\otimes|z\rangle$. 
Here 
\begin{equation}
{\mathcal N}_\pm(\alpha,z)=\sqrt{2}\left(1\pm e^{-2|\alpha|^2}\left(\frac{1-|z|^2}{1+|z|^2}\right)^{2j}\right)^{1/2}
\end{equation}
is a normalization 
factor. These even and odd coherent states are ``Schrodinger's cat states'' in the sense that they are a quantum 
superposition of quasi-classical, macroscopically distinguishable states. 
The new energy surface is now:
\begin{eqnarray}
 {\cal H}_\pm(\alpha,z)&=&\langle \alpha, z,\pm|H|\alpha, z,\pm\rangle\nonumber\\ 
 &=&\frac{{\cal H}(\alpha,z)\pm \langle \alpha,z|H|-\alpha,-z\rangle}{{\mathcal N}_\pm^2(\alpha,z)/2},
\end{eqnarray}
with non-diagonal elements
\begin{eqnarray} 
\langle \alpha,z|H|-\alpha,-z\rangle=e^{-2|\alpha|^2}\left(\frac{1-|z|^2}{1+|z|^2}\right)^{2j}\times \nonumber \\
\left(\omega|\alpha|^2-j\omega_0\frac{1+|z|^2}{1-|z|^2}+{\lambda}{\sqrt{2j}}(\alpha-\bar\alpha)\frac{z-\bar z}{1-|z|^2}\right)
\end{eqnarray}
The more involved structure of ${\cal H}_\pm(\alpha,z)$ makes much more difficult to obtain the new critical points 
$\alpha_e^{(\pm)},z_e^{(\pm)}$ minimizing the corresponding energy surface. Instead of carrying out a numerical computation of 
$\alpha_e^{(\pm)},z_e^{(\pm)}$ for different values of $j$ and $\lambda$, we shall use the approximation  
$\alpha_e^{(\pm)}\approx \alpha_e, z_e^{(\pm)}\approx z_e$, which turns out to be quite good 
except in a close neighborhood around $\lambda_c$ which diminishes as the
number of particles $N=2j$ increases (see Ref.
\cite{casta2}). 
With this approximation, we expect a rather good agreement between our numerical
and variational results except perhaps in a close vicinity of $\lambda_c$.

Taking into account the coherent state overlaps
\begin{eqnarray}
 \langle\alpha|\pm\alpha_e\rangle&=&e^{-\frac{1}{2}|\alpha|-\frac{1}{2}\alpha_e^2\pm\bar\alpha\alpha_e},\nonumber\\ \langle z|\pm z_e\rangle&=&\frac{(1\pm\bar zz_e)^{2j}}{(1+|z|^2)^j(1+z_e^2)^j},
\end{eqnarray}
the Husimi distribution for the variational states $|\alpha_e,z_e,\pm\rangle$ can be simply written as:
\begin{equation}
 {\Psi}_\pm(\alpha,z)=\frac{|\langle\alpha,z|\alpha_e,z_e\rangle\pm\langle\alpha,z|-
\alpha_e,-z_e\rangle|^2}{{\mathcal N}_\pm^2(\alpha_e,z_e)}.\label{Husimij}
\end{equation}

In order to compute the zeros, moments and entropies of ${\Psi}_\pm(\alpha,z)$, and to compare with numerical results of the previous section, 
we shall make use of the Holstein-Primakoff representation (\ref{hprep},\ref{hp1},\ref{contract}). 
With this approximation, the Husimi distribution can be cast in the new form
\begin{equation}
 {\Phi}_\pm(\alpha,\beta)=\frac{e^{-|\alpha|^2-|\beta|^2-\alpha_e^2-\beta_e^2}|e^{\bar\alpha\alpha_e+\bar\beta\beta_e}\pm 
e^{-\bar\alpha\alpha_e-\bar\beta\beta_e}|^2}{2\left(1\pm e^{-2\alpha_e^2-2\beta_e^2}\right)},\label{Husiminf}
\end{equation}
From now on we shall restrict ourselves to the even case and simply denote by 
$\psi=\psi_+$ and  ${\Phi}={\Phi}_+$ the wave function and the Husimi distribution of the variational ground state. 

\subsection{Moments and R\'enyi-Wehrl entropy of the Husimi distribution}

Important quantities to visualize the QPT in the Dicke model accross the critical point $\lambda_c$ will be the 
$\nu$th moments of the Husimi distrubution \eqref{momentsnu}. In particular, the inverse participation ratio is given by
\begin{equation}
P_j(\lambda)=M_{j,2}(\lambda)=\frac{1+\mathrm{sech}^2(\alpha_e^2+\beta_e^2)}{8}\label{partvar}
\end{equation}
\begin{figure}
\includegraphics[width=6cm]{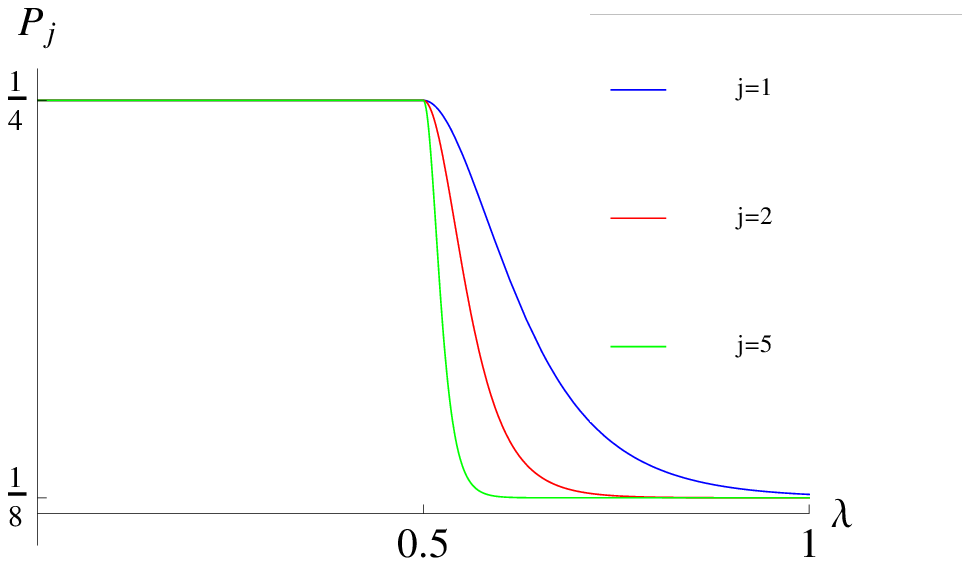}
\caption{Inverse participation ratio of the Husimi distribution as a function of $\lambda$ for different values of $j$ and $\lambda_c=0.5$.
\label{parvar}}
\end{figure}
Figure \ref{parvar} shows that $P_j(\lambda)$ tends to a Heaviside-type step function
\begin{equation}
P_\infty(\lambda)=\left\{\begin{array}{ll} 1/{4}, & \mathrm{if}\, \lambda<\lambda_c\\
{1}/{8}, &
\mathrm{if}\, \lambda\geq\lambda_c,\end{array}\right.
\end{equation}
which suffers a sudden decrease of $1/4$ from normal to superradiant phase, thus indicating a delocalization of 
$\Phi$ above the critical point $\lambda_c$. 
A similar behavior is displayed by higher moments in the thermodynamic limit
\begin{equation}
M_{j,\nu}(\lambda)\stackrel{j\to\infty}{\longrightarrow}\left\{\begin{array}{ll} {\nu^{-2}}, & \mathrm{if}\, \lambda<\lambda_c\\
{2^{1-\nu}\nu^{-2}}, &
\mathrm{if}\, \lambda\geq\lambda_c.\end{array}\right.\label{moments}
\end{equation}
The definition of the moments $M_{j,\nu}$ is not restricted to integer values of $\nu$.  Once $M_{j,\nu}$ are known for all integer $\nu$, 
there is a unique analytic extension to real $\nu>0$ (and to the right half complex plane).  This analytic extension is possible due to 
the particular expression of $\Phi$ in therms of Gaussian bells. Using \eqref{moments}, we can easily compute R\'enyi-Wehrl entropies \eqref{wehrlnu} and, in the limit $\nu\to 1$, the  
Wehrl entropy \eqref{wehrl1} in the thermodynamic limit
\begin{equation}
W_j(\lambda)\stackrel{j\to \infty}{\longrightarrow}\left\{\begin{array}{ll} 2, & \mathrm{if}\, \lambda<\lambda_c\\
2+\ln(2), &
\mathrm{if}\, \lambda\geq\lambda_c.\end{array}\right.
\end{equation}
This result is in agreement with the (still unproved) Lieb's conjecture. Indeed, as conjectured by Wehrl \cite{Wehrl} 
and proved by Lieb \cite{Lieb}, any Glauber coherent state $|\alpha\rangle$ has a minimum Wehrl entropy of 1. 
In the same paper by Lieb \cite{Lieb}, it was also conjectured that the extension of
Wehrl's definition of entropy for coherent spin-$j$ states will yield a minimum entropy $j/(j+1)$. For the joined 
system of radiation field plus atoms we would have $W_j(\lambda)=1+j/(j+1)$ in the normal phase ($\lambda<\lambda_c$), and therefore, 
$W_j\to 2$ in the thermodynamic limit, in agreement with our result. 

\subsection{Marginals of the Husimi distribution}

The explicit expression of the marginal Husimi distributions \eqref{marginalQ} for our variational ground state $\psi$ are
\begin{eqnarray}
\Phi_1(\alpha_1,\beta_1)&=&\frac{1+e^{\alpha_e^2+\beta_e^2}
\cosh(2(\alpha_1\alpha_e+\beta_1\beta_e))}{e^{\alpha_1^2+\beta_1^2}(1+e^{2\alpha_e^2+2\beta_e^2})},\nonumber\\ 
\Phi_2(\alpha_2,\beta_2)&=&\frac{1+e^{-\alpha_e^2-\beta_e^2}
\cos(2(\alpha_2\alpha_e+\beta_2\beta_e))}{e^{\alpha_2^2+\beta_2^2}(1+e^{-2\alpha_e^2-2\beta_e^2})}.
\end{eqnarray}

Using the definition \eqref{momentsnumarg}, we can compute inverse participation ratios for marginal distributions 
as a function of 
$\zeta_j(\lambda)\equiv e^{\alpha_e^2+\beta_e^2}$:
\begin{eqnarray}
P_j^{(1)}(\lambda)&=&\frac{2+4\zeta^{3/2}_j(\lambda)+\zeta^2_j(\lambda)+\zeta^4_j(\lambda)}{4(1+\zeta^2_j(\lambda))^2},\nonumber\\ 
P_j^{(2)}(\lambda)&=&\frac{1+\zeta^{2}_j(\lambda)+4\zeta^{5/2}_j(\lambda)+2\zeta^4_j(\lambda)}{4(1+\zeta^2_j(\lambda))^2}.\label{margpartvar}
\end{eqnarray}
\begin{figure}
\includegraphics[width=6cm]{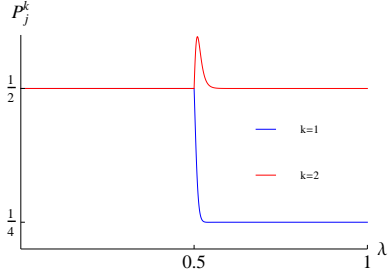}
\caption{Marginal inverse participation ratios $P_j^{(\kappa)}, \kappa=1,2$ (position and momentum, resp.)  
of the Husimi distribution as a function of 
$\lambda$ for $j=10$ and $\lambda_c=0.5$.
\label{parvarm}}
\end{figure}
Figure \ref{parvarm} shows $P_j^{(\kappa)}(\lambda)$ for $j=10$, indicating that $P_j^{(1)}(\lambda)$ suffers a 
sudden decrease from $1/2$ to $1/4$ across the phase transition, whereas $P_j^{(2)}(\lambda)$ remains constant 
(the small peak around $\lambda_c=0.5$ is perhaps an artifact due to the approximate character of $\alpha_e,\beta_e$ 
in a neighborhood of $\lambda_c$). In general, higher moments of marginal distributions in the thermodynamic limit are given by:
\begin{eqnarray}
M_{j,\nu}^{(1)}(\lambda)&\stackrel{j\to\infty}{\longrightarrow}&\left\{\begin{array}{ll} {\nu^{-1}}, & \mathrm{if}\, \lambda<\lambda_c\\
{2^{1-\nu}\nu^{-1}}, &
\mathrm{if}\, \lambda\geq\lambda_c.\end{array}\right.,\nonumber\\ 
M_{j,\nu}^{(2)}(\lambda)&\stackrel{j\to\infty}{\longrightarrow}& \nu^{-1}\,, \,\,\forall \lambda,
\label{momentsm}
\end{eqnarray}
so that, in this limit, we have $M_{j,\nu}(\lambda)=M_{j,\nu}^{(1)}(\lambda)M_{j,\nu}^{(2)}(\lambda)$.  
This equality is not true in general for finite $j$ 
and $\lambda>\lambda_c$, as can be directly checked for $\nu=2$ from \eqref{partvar} and \eqref{margpartvar}.  Now we see that the entropy excess 
of $\ln(2)$ comes from the position contribution, since Wehrl entropy in
momentum space remains constant:
\textcolor{black}{
\begin{equation}
W^{(1)}_j(\lambda)\stackrel{j\to \infty}{\longrightarrow}\left\{\begin{array}{ll} 1, & \mathrm{if}\, \lambda<\lambda_c\\
1+\ln(2), &
\mathrm{if}\, \lambda\geq\lambda_c.\end{array}\right.
\label{w1}
\end{equation}
\begin{equation}
W^{(2)}_j(\lambda)\stackrel{j\to \infty}{\longrightarrow}  1
\label{w2}
\end{equation}
}

In order to connect with (\ref{marginalposi}), we introduce 
position and momentum operators for the two bosonic modes as in (\ref{posmomenop}). Taking into account 
the position and momentum representation of an ordinary (canonical) CS  \eqref{csposmom}, 
the explicit expression of the ground
state wave function $|\alpha_e,\beta_e\rangle_+$ in position
($\psi(x,y)=\langle x,y|\alpha_e,\beta_e,+\rangle$) and momentum 
($\tilde\psi(p_x,p_y)=\langle p_x,p_y|\alpha_e,\beta_e,+\rangle$) representations can be
easily obtained as:
\begin{eqnarray}
\psi(x,y)&=&\sqrt{\frac{\omega\omega_0}{\pi}}{\mathcal N}_+(\alpha_e,\beta_e)\nonumber\\ & &\times
\left(e^{-\frac{1}{2}{(\sqrt{\omega}\,x-\sqrt{2}\,\alpha_e)^2}-\frac{1}{2}
{(\sqrt{\omega_0}\,y-\sqrt{2}\,\beta_e)^2}}\right.\label{wavefuncp}\\ & &+
\left.e^{-\frac{1}{2}{(\sqrt{\omega}\,x+\sqrt{2}\,\alpha_e)^2}-\frac{1}{2}{(\sqrt{\omega_0}\,y+\sqrt{2
}\,\beta_e)^2}}\right),\nonumber
\end{eqnarray}
\begin{eqnarray}
\tilde\psi(p_x,p_y)&=&\frac{2}{\sqrt{\omega\omega_0\pi}}{\mathcal N}_+(\alpha_e,\beta_e)\label{wavefuncm}\\ & &\times e^{-\frac{p_x^2}{2\omega}-\frac{p_y^2}{2\omega_0}}
\cos\left(\sqrt{2}(\frac{p_x}{\sqrt{\omega}}\alpha_e+\frac{p_y}{\sqrt{\omega_0}}\beta_e
)\right),\nonumber
\end{eqnarray}
where ${\mathcal N}_+(\alpha_e,\beta_e)=\left({2}(1+e^{-2\alpha_e^2-2\beta_e^2})\right)^{-1/2}$ 
is the typical  normalization factor obtained earlier.  Note that for $\lambda>\lambda_c$ the
ground-state wave function $\psi(x,y)$ splits up into two Gaussian
packets centered at antipodal points $\sqrt{2}(\alpha_e,\beta_e)$ and $-\sqrt{2}(\alpha_e,\beta_e)$ 
in the $x-y$ plane. The packets move away from
 each other for increasing $j$ above the critical point $\lambda>\lambda_c$. 
In momentum space, $\tilde\psi(p_x,p_y)$ is
 a Gaussian modulated by a cosine function which oscillates rapidly for high $j$
for $\lambda>\lambda_c$.

\subsection{Zeros of the Husimi distribution and {QPT}}

It is well known that the Husimi density is determined by its zeros through the Weierstrass-Hadamard factorization. 
It has also been observed that the distribution of zeros differs for classically regular or
chaotic systems and can be considered as a quantum indicator of classical chaos (see e.g. \cite{Korsch,monteiro,4}).

\begin{figure}
\includegraphics[width=4.5cm]{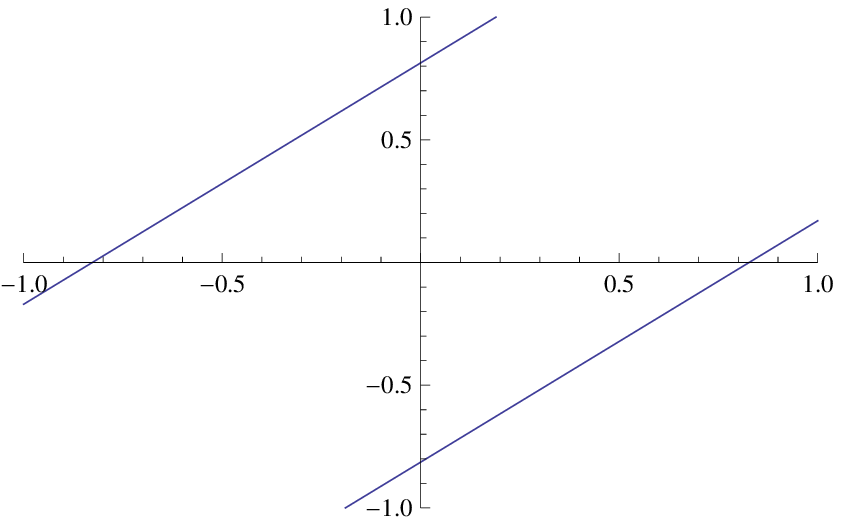}\includegraphics[width=4.5cm]{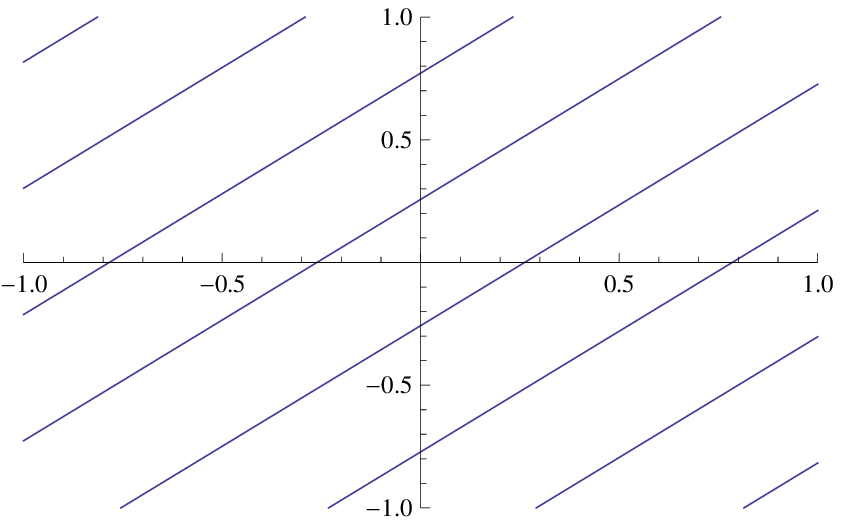}
\includegraphics[width=4.5cm]{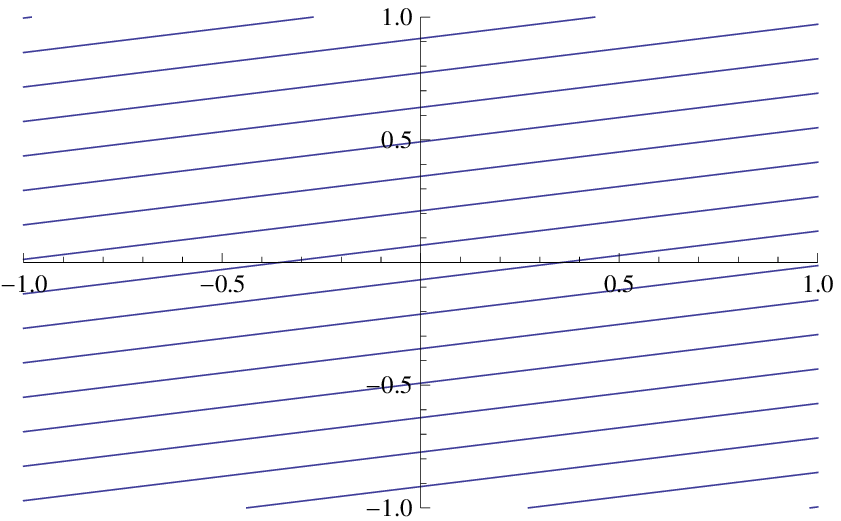}\includegraphics[width=4.5cm]{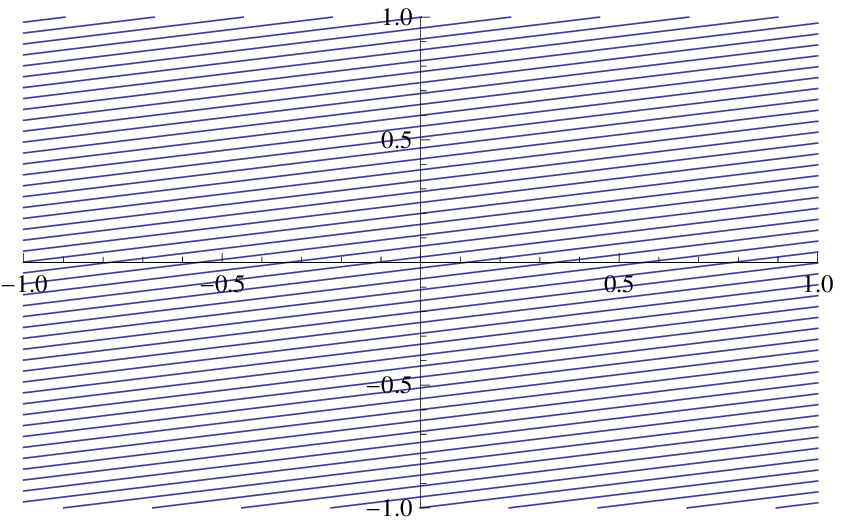}
\caption{Zeros of the Husimi distribution $\Phi(\alpha,\beta)$ in the cell $\alpha_2,\beta_2\in[-1,1]$ of the momentum plane 
for $\lambda=0.6, j=10$ (top-left),  $\lambda=0.6, j=100$ (top-right), $\lambda=10, j=10$ (bottom-left) and 
 $\lambda=10, j=100$ (bottom-right) for $\lambda_c=0.5$.
\label{zerosfig}}
\end{figure}

Here we shall explore the distribution of zeros of the Husimi density as a fingerprint 
of QPT in the Dicke model. From \eqref{Husimij} we obtain
\begin{equation}
 {\Psi}(\alpha,z)=0\Rightarrow 2\bar\alpha\alpha_e+2j\ln\frac{1+\bar z z_e}{1-\bar z z_e}=i\pi (2l+1),\,l\in\mathbb Z.
\end{equation}
Instead of this condition, we shall use the approximation \eqref{contract} and from \eqref{Husiminf} obtain
\begin{equation}
 {\Phi}(\alpha,\beta)=0\Rightarrow 2\bar\alpha\alpha_e+2\bar\beta\beta_e=i\pi (2l+1),\,l\in\mathbb Z,
\end{equation}
which is equivalent to
\begin{eqnarray}
 \alpha_1&=&-\frac{\beta_e}{\alpha_e}\beta_1,\label{zeros1}\\
 \alpha_2&=&-\frac{\beta_e}{\alpha_e}\beta_2-\frac{\pi}{2\alpha_e}(2l+1).\label{zeros2}
\end{eqnarray}
We see that, in the normal phase ($\alpha_e=0=\beta_e$) the Husimi distribution $\Phi(\alpha,\beta)$ has no zeros. In the 
superradiant phase ($\lambda>\lambda_c$) the zeros are localized along straight lines (``dark fringes'') in the 
$\alpha_1\beta_1$ (position) and $\alpha_2\beta_2$ (momentum) planes. In the momentum plane, the number of dark 
fringes per cell $\alpha_2,\beta_2\in[-1,1]$ grows with $\lambda$ and $j$, as depicted in Figure \ref{zerosfig}. In the 
thermodynamic limit $j\to\infty$, zeros densely fill the momentum plane $\alpha_2\beta_2$. 

\section{Conclusions}\label{sec3}

We have found that inverse participation ratios and R\'enyi-Wehrl entropies of the Husimi distribution provide 
sharp indicators of a quantum phase transition in the Dicke model. This uncertainty measures detect a delocalization of the 
Husimi distribution across the critical point $\lambda_c$ and we have employed them to quantify the phase-space spreading of
the states. 
The advantage of working in phase space is that 
we can analyze contributions in position and momentum space jointly and separately. Marginal magnitudes in position space 
turn out to provide sharper indicators of the QPT than in momentum space, where these quantities remain nearly constant. 
However, zeros of the Husimi distribution exhibit a richer structure in momentum than in position space. 

Calculations have been done numerically and through a variational approximation. Numerical calculations are 
performed by using  explicit 
expressions which have been  derived by adopting  a truncation of the Holstein-Primakoff
representation of the angular momentum operators. The variational approach, in terms of symmetry-adapted 
coherent states, complements and 
enriches the analysis  providing explicit analytical
expressions for the inverse participation ratios and R\'enyi-Wehrl entropies which remarkably coincide with the
numerical results, especially in the thermodynamic limit.

In the superradiant phase, 
Wehrl's entropy undergoes an 
entropy excess (or ``subentropy'' \cite{Jozsa}) of $\ln(2)$. This fact implies that the Husimi distribution splits 
up into two identical subpackets with negligible overlap in passing from normal to superradiant phase. In general, for 
$s$ identical subpackets with negligible
overlap, one would expect an entropy excess of $\ln(s)$. We would like to mention that the Wehrl subentropy (or excess of the Wehrl
entropy)  has been also used in \cite{Mintert} as a measure of the degree of mixing for monopartite states or 
of the degree of entanglement for pure states of bipartite systems. 
Dicke model is also known to exhibit entanglement between the atoms and the field \cite{emaryprl2,emarypra} and 
a characterization of it in terms of entropy excesses of this kind would be interesting.

The QPT fingerprints in the Dicke model have also been tracked by exploring
the distribution of zeros of the Husimi density {within the
  analytical variational approximation. We have found that the zeros
  characterize the QPT in this model. Moreover,    zeros densely fill the
  momentum plane  in the superradiant phase for the ground state variational approximation in the thermodynamic limit}. 
 This subject 
deserves further attention and should be studied in other models too. For the moment, we have detected a sudden growth of zeros above the critical point 
$\lambda_c$ in the Holstein-Primakoff approximation.

\section*{Acknowledgments}

This work was supported by the Projects:   FIS2011-24149 and FIS2011-29813-C02-01 (Spanish MICINN),  
FQM-165/0207 and FQM219 (Junta de Andaluc\'\i a). 

\begin{appendix}
\section{Marginal Husimi distributions as Gaussian smearings\label{apA}}
Working on resonance $\omega=\omega_0$, we can introduce  a ``natural variance'' $\sigma^2=1/(2\omega)$ 
in the Dicke model by considering the change of coordinates 
\begin{equation}\begin{array}{ll}
\alpha_1=\frac{x}{2\sigma}, &  \alpha_2=\sigma k_x,\\
\beta_1=\frac{y}{2\sigma}, &  \beta_2=\sigma k_y,\label{notation}
\end{array}
\end{equation}
with $\mathbf{r}=(x,y), \mathbf{k}=(k_x,k_y)$ position and momentum vectors.   Taking into account 
the position and momentum representation of an ordinary (canonical) CS  \cite{Perelomov}:
\begin{eqnarray}
 \langle x|\alpha\rangle&=&\left(\frac{\omega^2}{\pi}\right)^{1/4}e^{i\sqrt{2\omega}\,\alpha_2 x}
e^{-{(\sqrt{\omega}\,x-\sqrt{2}\,\alpha_1)^2}/{2}},\nonumber\\
 \langle k|\alpha\rangle&=&\left(\frac{1}{\pi\omega^2}\right)^{1/4}e^{i\sqrt{\frac{2}{\omega}}\,\alpha_1 k}
e^{-{(\frac{k}{\sqrt{\omega}}-\sqrt{2}\,\alpha_2)^2}/{2}},\label{csposmom}
\end{eqnarray}
one can easily see that marginal distributions \eqref{marginalQ} can be also obtained as a smearing (convolution  product) 
of the density functions $|\psi(\mathbf{r})|^2$ and $|\tilde{\psi}(\mathbf{k})|^2$ by Gaussians 
$g_{\sigma}({\bf r})=(2\pi\sigma^2)^{-1}\exp{\left(-{{\bf
      r}^2}/({2\sigma^2})\right)}$ and $\tilde{g}_{\sigma}({\bf k})=4\pi\sigma^2\exp{\left(-2\sigma^2{{\bf
      k}^2}\right)}$ as:
\begin{eqnarray}
 \xi({\bf r})&=&
\int d{\bf r}^{\prime} g_{\sigma}({\bf r}-{\bf r}^{\prime})
|\psi({\bf r}^{\prime})|^2
\label{marginalposi}\\
\tilde\xi({\bf k})&=&\int \frac{d{\bf k}^{\prime}}{(2\pi)^2} \tilde{g}_{\sigma}({\bf k}-{\bf k}^{\prime})
|\tilde{\psi}({\bf k}^{\prime})|^2
\nonumber
\end{eqnarray}
with $\int d{\bf r} \xi({\bf r})=1$ and $\int \frac{d{\bf k}}{(2\pi)^2}
\tilde\xi({\bf k})=1$. Inverse participation ratios and
and Wehrl entropies for these marginal distributions are now written as: 
\begin{equation}
P_j^{\xi}=\int d{\bf r} \xi^2({\bf r}), \quad P_j^{\tilde{\xi}}=\int \frac{d{\bf k}}{(2\pi)^2}
\tilde\xi^2({\bf k})
\end{equation}
\begin{equation}
W_j^{\xi}=\int d{\bf r} \xi({\bf r})\ln \xi({\bf r}), \quad W_j^{\tilde{\xi}}=\int \frac{d{\bf k}}{(2\pi)^2}
\tilde\xi({\bf k})\ln\tilde\xi({\bf k}).
\end{equation}
More, explicitly, from \eqref{wf} and \eqref{wfmomentum}, marginal Husimi distributions
(\ref{marginalposi}) are given in terms of the coefficients
$c_{nm}^{(j)}$ as:
\begin{equation}
\xi(x,y)= \sum_{n,m,n^{\prime},m^{\prime}} c_{nm}^{(j)}
c_{n^{\prime}m^{\prime}}^{(j)} I_{n ,n^{\prime}}(x)I_{m+j, m^{\prime}+j}(y)
\end{equation}
and 
\begin{equation}
\tilde\xi(k_x,k_y)= (2\pi)^2\sum_{n,m,n^{\prime},m^{\prime}}  d_{nm}^{(j)}
d_{n^{\prime}m^{\prime}}^{(j)} I_{n ,n^{\prime}}(p_x)I_{m+j, m^{\prime}+j}(p_y)
\label{eta}
\end{equation}
with $d_{nm}^{(j)}\equiv(-i)^{n+m+j}c_{nm}^{(j)}$ and 
\begin{eqnarray}
&& I_{n,n^{\prime}}(x)=A\alpha\sqrt{n! n^{\prime}!\left(\frac{1-\alpha^2}{2}\right)^{n+n^{\prime}}}
e^{\frac{-x^2}{1+2\sigma^2}}
\\
&&\times\sum_{k=0}^{\mu}
B(n,n^{\prime},k)
\left(\frac{2}{1-\alpha^2}\right)^kH_{n+n^{\prime}-2k}\left(\frac{\alpha
s}{(1-\alpha^2)^{1/2}}\right)
\notag
\end{eqnarray}
with  $\mu=\min\left\{n,n^{\prime}\right\}$, $A=(2\pi\sigma^2)^{-1/2}$, $B(n,n^{\prime},k)=\frac{1
}{(k!)(n-k)!(n^{\prime}-k)!}$,
 $\alpha=\sqrt{\frac{2\sigma^2}{2\sigma^2+1}}$ and 
$s=\frac{x}{\sigma\sqrt{2(1+2\sigma^2)}}$. Relations between marginal inverse participation ratios and 
Wehrl entropies in both cases can be straightforwardly obtained: 
$P_j^{\xi}=\pi^{-1}P_j^{(1)}$ and
$P_j^{\tilde{\xi}}=\pi P_j^{(2)}$ and $W_j^{\xi}=W_j^{(1)}+\ln(2\pi)$ and $W_j^{\tilde{\xi}}=W_j^{(2)}-\ln(2\pi)$. 

\end{appendix}


\begin{thebibliography}{100}

\bibitem{sachdev} S. Sachdev, {\em Quantum Phase Transitions}, Cambridge University Press
 (2000).

\bibitem{Gerry2005}C. C. Gerry and P. L. Knight, {\em Introductory Quantum
  Optics}, Cambridge University Press (2005).
\bibitem{aulbach}C. Aulbach, A. Wobst, G.-L. Ingold, P. Hanggi and I. Varga, New J. of
  Physics {\bf 6}, 70 (2004).
\bibitem{monteiro} P. A. Dando and T. S. Monteiro, J. Phys. B. {\bf 27}, 2681, (1994).
\bibitem{weinmann99}
D. Weinmann , S.  Kohler, G-L. Ingold  and P. H�nggi, Ann. Phys. (Lpz) 8 SI277 (1999).


\bibitem{Korsch} H. J. Korsch, C. M\"uller and H. Wiescher, J. Phys. A {\bf
  30}, L677 (1997).
\bibitem{4} P. Leboeuf and A. Voros, J. Phys. A {\bf 23}, 1765 (1990).
\bibitem{dando} P. A. Dando, and T. S. Monteiro, J. Phys. B {\bf 27}, 2681 (1994).
\bibitem{Mintert} F. Mintert and K. Zyczkowski, Phys. Rev. A \textbf{69},
  022317 (2004).
\bibitem{pla11} E. Romera and \'A. Nagy, Phys. Lett. A {\bf 375} 3066 (2011).
\bibitem{jsm11}E. Romera, K. Sen and \'A. Nagy, 
 J.Stat. Mech.
doi:10.1088/1742-5468/2011/09/P09016
\bibitem{epl2012} E. Romera, M. Calixto and \'A. Nagy, Europhys. Lett. 
{\bf 97}, 20011 (2012). 
\bibitem{physica12} \'A. Nagy and E. Romera, Physica A doi: 10.1016/j.physa.2012.02.024 (2012).
\bibitem{renyipra} M. Calixto, \'A. Nagy, I. Paraleda and E. Romera
  (submitted, 2012).
 \bibitem{emaryprl2} N. Lambert, C. Emary, and T. Brandes, Phys. Rev. Lett.
{\bf 92}, 073602 (2004). 
\bibitem{emarypra} N. Lambert, C. Emary, and T. Brandes, Phys. Rev.A \textbf{71}, 053804 (2005).
\bibitem{brandes} T. Brandes, Phys. Rep. {\bf 408}, 315 (2005).
\bibitem{Perelomov} A. Perelomov, Generalized Coherent States and Their
Applications, Springer-Verlag (1986).
\bibitem{HP} T. Holstein and H. Primakoff, Phys. Rev. {\bf 58}, 1098 (1940).
\bibitem{emarypre} C. Emary and T. Brandes, Phys. Rev. E {\bf 67}, 066203   (2003).
\bibitem{Hirchdetallado} M. A. Bastarrachea-Magnani, J. G. Hirsch
{\em Numerical solutions of the Dicke Hamiltonian}, arxiv:1108.0703 (2011)
\bibitem{CScontract1} J.M. Radcliffe, J. Phys. A {\bf 4}, 313 (1971).
\bibitem{casta1} O. Casta\~nos, E. Nahmad-Achar, R. L\'opez-Pe\~na and
  J. G. Hirsch, Phys. Rev. A {\bf 83}, 051601 (2011).
\bibitem{casta2} O. Casta\~nos, E. Nahmad-Achar, R. L\'opez-Pe\~na, and
J. G. Hirsch, Phys. Rev. A, {\bf 84} 013819 (2011).
\bibitem{Wehrl} A. Wehrl,  Rep. Math. Phys. \textbf{16}, 353 (1979)
\bibitem{Lieb}  E.H. Lieb, Commun. Math. Phys. \textbf{62}, 35 (1978)
\bibitem{Jozsa} R. Jozsa, D. Robb and W.K. Wootters, Phys. Rev. A \textbf{49}, 668 (1994).


\end{thebibliography}
\end{document}